\documentclass[%
prd,%
preprintnumbers,%
amsmath,%
amssymb,%
nofootinbib,%
twocolumn
]{revtex4}
\usepackage[utf8]{inputenc}

\usepackage{graphicx}
\usepackage{color}
\usepackage[dvipsnames]{xcolor}
\usepackage{epstopdf}
\usepackage{hyperref}
\usepackage{cancel} 

\newcommand{\bea}{\begin{eqnarray}}
\newcommand{\eea}{\end{eqnarray}}

\newcommand{\eq}[1]{Eq.~\eqref{#1}}

\newcommand{\noi}{\noindent}

\begin{document}
\preprint{CERN-TH-2021-092, LAPTH-020/21, PSI-PR-21-12, ZU-TH 27/21}

\title{\boldmath The Forward-Backward Asymmetry in $B\to D^{*}\ell\nu$:\\One more hint for Scalar Leptoquarks?}

\author{Alexandre Carvunis}
\email{alexandre.carvunis@lapth.cnrs.fr}
\affiliation{LAPTh, CNRS et Universit\'{e} Savoie Mont-Blanc, Annecy, France}

\author{Shireen Gangal}
\email{shireen.gangal@lapth.cnrs.fr}
\affiliation{LAPTh, CNRS et Universit\'{e} Savoie Mont-Blanc, Annecy, France}

\author{Andreas Crivellin}
\email{andreas.crivellin@cern.ch}
\affiliation{Department of Theoretical Physics, CERN, Geneva, Switzerland}
\affiliation{Physik-Institut, Universit\"at Z\"urich, Winterthurerstrasse 190, CH--8057 Z\"urich, Switzerland}
\affiliation{Paul Scherrer Institut, CH--5232 Villigen PSI, Switzerland}

\author{Diego Guadagnoli}
\email{diego.guadagnoli@lapth.cnrs.fr}
\affiliation{Department of Theoretical Physics, CERN, Geneva, Switzerland}
\affiliation{LAPTh, CNRS et Universit\'{e} Savoie Mont-Blanc, Annecy, France}

\begin{abstract}
Experimental data have provided intriguing hints for the violation of lepton flavour universality (LFU), including $B\to D^{(*)}\tau\nu/B\to D^{(*)}\ell\nu$, the anomalous magnetic moment of the muon and $b\!\to\! s\ell^+\ell^-$ with a significance of $\!>3\,\sigma$, $>\!4\,\sigma$ and $>\!5\,\sigma$, respectively.
Furthermore, in a recent re-analysis of 2018 Belle data, it was found that the forward-backward asymmetry ($\Delta A_{\rm FB}$) of $B \to D^{*}\mu\bar \nu$ vs $B\to D^{*}e\bar \nu$ disagrees with the SM prediction by $\approx\!\!4\,\sigma$, providing an additional sign of LFU violation.
We show that a tensor operator is necessary to significantly improve the agreement with data in $\Delta A_{\rm FB}$ while respecting the bounds from other $b\to c\ell\nu$ observables. Importantly, this tensor operator can only be induced (at tree-level within renormalizable models) by a scalar leptoquark. Furthermore, among the two possible representations, the $SU(2)_L$-singlet $S_1$ and the doublet $S_2$, which can interestingly both also account for the anomalous magnetic moment of the muon, only $S_1$ can provide a good fit. Even though the constraints from (differences of) other angular observables prefer a smaller value of $\Delta A_{\rm FB}$ than the current central one, this scenario is significantly preferred (nearly $4 \sigma$) over the SM hypothesis, and is compatible with constraints such as $B\to K^*\nu\nu$ and electroweak precision bounds. Therefore, if the $\Delta A_{\rm FB}$ anomaly is confirmed, it would provide circumstantial evidence for scalar leptoquarks and pave the way for a natural connection with all other anomalies pointing towards LFU violation.
\end{abstract}


\maketitle

\section{Introduction}

\noi In the last decades, the Standard Model (SM) of particle physics has been extensively tested and verified, both in high-energy searches and in low-energy precision experiments. The absence of any direct evidence of a new particle at the LHC, and the striking success of the SM in the vast majority of measurements nearly drowned the fact that it cannot be the ultimate theory of physics as it e.g. does not account for Dark Matter or neutrino masses. Therefore, the SM should be thought of as the dim$\le 4$ part of an effective theory where higher dimensional operators parametrize the effect of new physics~\cite{Appelquist:1974tg}.

In fact, signs for the presence of non-vanishing Wilson coefficients of such higher dimensional operators have accumulated within recent years in flavour observables. In particular, $b\!\to\! s\ell^+\ell^-$ data~\cite{Aaij:2017vbb,Aaij:2019wad,Aaij:2020nrf}, $b\!\to\! c\tau\nu$ transitions~\cite{Lees:2012xj,Aaij:2017uff,Abdesselam:2019dgh} and the anomalous magnetic moment (AMM) of the muon ($a_\mu=(g-2)_{\mu}/2$)~\cite{Bennett:2006fi,Mohr:2015ccw,Abi:2021gix} show deviations from their SM predictions with a significance of $>\!5\,\sigma$~\cite{Capdevila:2017bsm,Altmannshofer:2017yso,DAmico:2017mtc,Ciuchini:2017mik,Hiller:2017bzc,Geng:2017svp,Hurth:2017hxg,Alok:2017sui,Alguero:2019ptt,Aebischer:2019mlg,Ciuchini:2019usw,Ciuchini:2020gvn,Carvunis:2021jga,Hurth:2021nsi,Altmannshofer:2021qrr,Alguero:2021anc}, $>\!3\,\sigma$~\cite{Amhis:2016xyh,Murgui:2019czp,Shi:2019gxi,Blanke:2019qrx,Kumbhakar:2019avh} and $4.2\,\sigma$~\cite{Aoyama:2020ynm}, respectively.

The observation that all these deviations are instances of lepton flavour universality violation (LFUV) \footnote{Interestingly, also the Cabibbo Angle Anomaly~\cite{Belfatto:2019swo,Grossman:2019bzp,Seng:2020wjq} and the CMS excess in non-resonant electron pairs~\cite{Sirunyan:2021khd} can be understood in this context~\cite{Coutinho:2019aiy,Crivellin:2020lzu,Crivellin:2021rbf,Kirk:2020wdk,Capdevila:2020rrl}.} suggests that they could (or even should) be related. Under this assumption, one might naturally expect new-physics effects in $b\!\to\! c\mu\nu / b\!\to \!c e\nu$ ratios~\cite{Jung:2018lfu} as well, since it is a $b\!\to\! c$ transition, and like the aforementioned observables it involves muons. Even though the corresponding measurements of the total branching ratios are consistent with the SM expectations~\cite{Glattauer:2015teq,Abdesselam:2017kjf}, recently Ref.~\cite{Bobeth:2021lya} unveiled a $\approx\!4\sigma$ tension in the difference of the forward-backward asymmetries, $\Delta A_{\mathrm{FB}}\equiv A_{\mathrm{FB}}^{(\mu)}-A_{\mathrm{FB}}^{(e)}$, extracted from $B\to D^*\ell\bar \nu$ data in Ref.~\cite{Waheed:2018djm}\footnote{Ref.~\cite{Bobeth:2021lya} notes that the presentation of the data in Ref.~\cite{Waheed:2018djm} does not allow for a reliable determination of the correlation between the systematic uncertainties in $A^{\mu}_{\rm FB}$ and in $A^e_{\rm FB}$. However, adopting the most conservative assumption of $\rho = -1$ leads to a $\Delta A_{\mathrm{FB}}$ discrepancy of 3.6$\sigma$, such that varying $\rho$ from $\rho = -1$ (deemed as unrealistic) the discrepancy will only increase.}.

Awaiting further scrutiny, this additional hint for LFUV is especially interesting as $b\to s\ell^+\ell^-$ data and $a_\mu$ clearly suggest new physics related to muons. Therefore one naturally expects muon-related new effects to also emerge at some point in $b \to c\mu\nu$ decays, and $\Delta A_{\mathrm{FB}}$ might be a first manifestation. Furthermore, a sizeable deviation from the SM prediction in the forward-backward asymmetry requires scalar and/or tensor operators, naturally relating $\Delta A_{\mathrm{FB}}$ to $a_\mu$, which is a chirality-violating observable as well \footnote{Since $S_1$ gives rise to scalar and tensor operators, it can also produce interesting effects in electric dipole moments~\cite{Crivellin:2018qmi,Fuyuto:2018scm,Dekens:2018bci,Crivellin:2019qnh,Altmannshofer:2020ywf}.}.

In particular, as we will see, a tensor operator is required to significantly improve the SM fit. At tree level, within renormalizable extensions of the SM, of all possible scalar, spinor or vector mediators, only two representations of scalar leptoquarks (LQs) can generate such tensor operators \cite{Crivellin:2017dsk,deBlas:2017xtg} and only one of them ($S_1$) yields the right linear combinations of Wilson coefficients to neatly describe the effect in $\Delta A_{\rm FB}$. This $SU(2)_L$-singlet scalar LQ $S_1 \sim \left( {\bf 3}, {\bf 1}, - 1/3\right)$ is well-motivated, since it is e.g.~present within the R-parity violating MSSM in the form of the right-handed down-squark~\cite{Hall:1983id,Ross:1984yg,Barger:1989rk,Dreiner:1997uz,Barbier:2004ez}\footnote{Note that in the minimal R-parity violating MSSM the coupling to charge-conjugated fields in \eq{LLQ} is absent. For an analysis of EDM constraints within this setup see Ref.~\cite{Yamanaka:2014nba}.}. Besides, as well-known, it provides a possible explanation of $a_\mu$ and $b\!\to\! c\tau\nu$ data: it can account for the former by an $m_t/m_\mu$-enhanced contribution~\cite{Bauer:2015knc,Djouadi:1989md, Chakraverty:2001yg,Cheung:2001ip,Popov:2016fzr,Chen:2016dip,Biggio:2016wyy,Davidson:1993qk,
Couture:1995he,Mahanta:2001yc,Queiroz:2014pra, ColuccioLeskow:2016dox,Becirevic:2016oho,
Chen:2017hir,Das:2016vkr,Cai:2017wry,Crivellin:2018qmi,Kowalska:2018ulj,Dorsner:2019itg,Crivellin:2019dwb,DelleRose:2020qak,Saad:2020ihm,Bigaran:2020jil,Dorsner:2020aaz,Fuentes-Martin:2020bnh,Babu:2020hun,Crivellin:2020tsz,Greljo:2021xmg,Zhang:2021dgl} and enters $b\to c\tau\nu$ processes at tree level~\cite{Fajfer:2012jt,Deshpande:2012rr,Tanaka:2012nw,Sakaki:2013bfa,Freytsis:2015qca,Hati:2015awg,Bauer:2015knc,Li:2016vvp,Zhu:2016xdg,Popov:2016fzr,Deshpand:2016cpw,Crivellin:2017zlb,Altmannshofer:2017poe,Kamali:2018fhr,Azatov:2018knx,Wei:2018vmk,Hu:2018lmk,Angelescu:2018tyl,Kim:2018oih,Yan:2019hpm,Crivellin:2019dwb,Angelescu:2021lln} where it gives a very good fit to data (including polarization observables)~\cite{Feruglio:2018fxo,Iguro:2018vqb,Blanke:2018yud,Murgui:2019czp} since it generates vector, scalar and tensor operators. Finally, slight extensions of the model also allow for a combined explanation of $b\!\to\! c\tau\nu$ and $b\!\to\! s\ell^+\ell^-$ data~\cite{Crivellin:2017zlb,Bigaran:2019bqv,Gherardi:2020qhc,Crivellin:2019dwb,DaRold:2020bib,Marzocca:2021azj,Greljo:2021xmg,Altmannshofer:2020axr}. In particular, the singlet-triplet LQ model~\cite{Crivellin:2017zlb}, containing $S_1$ and $S_3$ can account for all three anomalies~\cite{Crivellin:2019dwb,Gherardi:2020qhc} and we will return to this model at end of our analysis.
These arguments make $\Delta A_{\rm FB}$, and related observables, a potential turning point in the understanding of the anomalies in muonic data as they could establish the existence of scalar LQs.

\section{Setup and Observables}
\label{observables}

In this section we establish our setup, calculate the predictions for the relevant observables and discuss their current experimental status and future prospects. 

\subsection{\boldmath The Scalar $SU(2)_L$-Single LQ}

The scalar leptoquark $S_1 \sim \left( {\bf 3}, {\bf 1}, - 1/3\right)$ couples to SM fermions via the Lagrangian
\begin{align}
	\mathcal{L} =& \left( \lambda_{fi}^{L}\overline {Q_f^c} i{\tau _2}{L_i}  +{\lambda}^{R}_{fi}\overline{u^c_f}\ell_i\right)S_1^{\dagger} + {\rm{h}}{\rm{.c}}.\,.\label{LLQ}
\end{align}
Here, $L$ ($Q^c$) is the lepton (charge-conjugated quark) $SU(2)_L$ doublet,  $\ell$ ($u^c$) the charged lepton (charge-conjugated up quark) singlet and $f,i$ are flavor indices. 

After electro-weak symmetry breaking, the Lagrangian in~\eq{LLQ} decomposes into components of definite electric charge. Absorbing unphysical rotations from the diagonalization of the fermion mass matrices into the couplings $\lambda_{fi}^{L,R}$ we have
\[{\cal L}_{\rm eff}^{\rm EW} \!=\! \left( {\lambda _{fi}^R\bar u_f^c{P_R}{\ell _i}\! +\! V_{fj}^*\lambda _{ji}^L\bar u_f^c{P_L}{\ell _i}\! -\! \lambda _{fi}^L\bar d_f^c{P_L}{\nu _i}} \right)\!S_1^\dag  + {\rm{h}}.{\rm{c}}.,\]
in the down-basis where the CKM matrix $V$ appears in the couplings to left-handed up-type quarks. We denote the mass of the LQ by $M$ and neglect its couplings to the SM Higgs which are expected to have small phenomenological consequences. The most relevant classes of observables in our model are $b\to s\nu\nu$ and $b\to c\ell\nu$ transitions,  as well as modified $Z$-$\mu\mu$ and $W$-$\mu\nu$ couplings.

\subsection{\boldmath $b\to s\nu \nu$}

For $b\to s\nu \nu$ transitions we follow the conventions of Ref.~\cite{Buras:2014fpa}
\begin{align}
&\mathcal{H}_{{\rm eff}}^{\nu\nu}=-\dfrac{4G_F}{\sqrt{2}}V_{td_k}V_{td_j}^{*} \left(C^{fi}_{L,jk}\mathcal{O}_{L,jk}^{fi}+C^{fi}_{R,jk}\mathcal{O}_{R,jk}^{fi}\right)\,,
\nonumber\\
&\mathcal{O}_{L(R),jk}^{fi}=\frac{\alpha}{4\pi}\left[\bar{d}_{j} \gamma^{\mu}P_{L(R)}d_{k}\right] \left[\bar{\nu}_{f}\gamma_{\mu}\left(1-\gamma_5\right)\nu_i\right]\,,
\end{align}
and obtain, already at tree level, the contribution
\begin{align}
C_{L,jk}^{fi\,{\rm NP}} = - \frac{{\sqrt 2 }}{{4{G_F}{V_{t{d_k}}}V_{t{d_j}}^*}}\frac{\pi }{\alpha }\frac{{\lambda _{jf}^{L*}\lambda _{ki}^L}}{{M_{}^2}}\,,\label{bsnunu}
\end{align}
with $C_{L,sb}^{{\rm SM},fi}\approx-1.47/s_W^2\delta_{fi}$. The best constraint originates from $B\to K^{(*)}\nu \bar\nu$, whose branching ratios, normalized to the corresponding SM predictions, read
\begin{equation}
{R_{K^{(*)}}^{\nu\bar{\nu}}} = 
\frac{1}{3}\sum\limits_{f,i=1}^3 \dfrac{ \big|{C_{L,sb}^{fi}}\big|^2}{\big|{C_{L,sb}^{{\rm SM},ii}}\big|^2} \,,
\end{equation}
for which the current experimental limits are ${R_K^{\nu\bar{\nu}}} < 3.9$ and ${R_{{K^*}}^{\nu\bar{\nu}}} < 2.7$~\cite{Grygier:2017tzo} (both at $90\%\,\mathrm{C.L.}$). The future BELLE II sensitivity for $B\to K^{(*)}\nu\bar{\nu}$ is 30\% of the SM branching ratio~\cite{Abe:2010gxa}.

\subsection{\boldmath $b\to c\ell\nu$}

For charged-current leptonic $b\to c\ell\nu$ decays we define the weak-effective-theory (WET) Hamiltonian as 
%
\begin{align}
{\cal H}_{{\rm{eff}}}^{\ell \nu } = \frac{{4{G_F}}}{{\sqrt 2 }}{V_{{c}b}}\left( {C_{VL}^\ell O_{VL}^\ell + C_{SL}^\ell O_{SL}^\ell + C_T^\ell O_T^\ell} \right)\,,\nonumber
\end{align}
with the operators given by
\begin{align}
\begin{aligned}
O_{VL}^{\ell} &= \bar c \gamma^\mu P_L b \,\,\, \bar \ell \gamma_\mu P_L \nu_\ell \,,\\
O_{SL}^{\ell} &= \bar c P_L b \,\,\, \bar \ell P_L \nu_\ell\,,\\
O_T^{\ell} &= \bar c \sigma^{\mu \nu} P_L b \,\,\, \bar \ell \sigma_{\mu \nu} P_L \nu _\ell \,.
\end{aligned}
\label{operators}
\end{align} 
In the SM $C_{VL}^{\ell}=1$ and we only included lepton flavour conserving operators that can interfere with the SM. The matching of our $S_1$ LQ model on this effective Hamiltonian gives
\begin{align}
\begin{aligned}
C_{VL}^\ell &= \frac{{\sqrt 2 }}{{8{G_F} \, V_{cb}}}\frac{V_{{c}j}{\lambda _{j\ell}^{L*}\lambda _{3\ell}^L}}{{M_{}^2}}\,, \\
C_{SL}^\ell &=  - 4C_T^\ell = -\frac{{\sqrt 2 }}{{8{G_F}V_{cb}}}\frac{{\lambda _{2\ell}^{R*}\lambda _{3\ell}^L}}{{M_{}^2}}\,.
\end{aligned}
\end{align}
These matching conditions can be improved by including QCD effects to one loop \cite{Aebischer:2018acj} such that the two-loop QCD RGE for the scalar and tensor operators~\cite{Gracey:2000am,Gonzalez-Alonso:2017iyc} can be taken consistently into account. Numerically, this RGE evolution is given by
\begin{align}
\left( \begin{array}{c}
C_{SL}^{{ \ell}}(m_b) \\ C_T^{{ \ell}}(m_b)           
\end{array}\right) \approx  \left( \begin{array}{rr}
1.9 & -0.36 \\
0 & 0.90 
\end{array}\right)
\left( \begin{array}{c}
C_{SL}^{{ \ell}}({M}) \\ C_{T}^{{ \ell}}({M})    
\end{array}\right) ,
\end{align}
for a matching scale $M =$ 1.5~TeV. Note that this includes the one-loop EW RGE, even though the operators are not manifestly $SU(2)_L$-invariant. Importantly, at the $B$ mass scale this implies the relation $C^\mu_{SL} \approx -8.5 \, C^\mu_T$, which is also preferred by the global fit as we will see later.

With these Wilson coefficients we can now calculate the resulting $b\to c\mu\nu$ observables including 
\begin{equation}
R(D^{(*)})_{\mu e}={\rm Br}[B\to D^{(*)}\mu\nu]/{\rm Br}[B\to D^{(*)}e\nu]
\end{equation}
measured by BELLE ~\cite{Glattauer:2015teq,Abdesselam:2017kjf,Waheed:2018djm} and the full single-differential lepton-averaged distributions of $\bar B \to D^{*} \ell \bar \nu$ ($\ell = \mu,e$)~\cite{Abdesselam:2017kjf}. The same data was later reanalysed to extract lepton-specific quantities for both $\ell = \mu, e$ \cite{Waheed:2018djm}. We include in our global fit the most constraining observables extracted in~\cite{Bobeth:2021lya} from~\cite{Waheed:2018djm}, namely $\Delta A_{\rm FB},\Delta S_3,\Delta F_L$ and $\Delta \tilde F_L$ and their correlation. The most important among these observables read
	\begin{equation}
	\begin{aligned}
	R(D_{\mu e})^{\rm exp}&=0.995 \pm 0.022 \pm 0.039\,,\\
	R(D^*_{\mu e})^{\rm exp}&=0.99 \pm 0.01 \pm 0.03\,,\\
	\Delta A_{\rm FB}^{\rm exp}&\approx0.035 \pm 0.009\,,\\
	\Delta S_3^{\rm exp}& \approx -0.013 \pm 0.01 \,,\\
	\Delta F_L^{\rm exp}&= -0.0065 \pm 0.0059 \,,\\
	\Delta \tilde F_L^{\rm exp} & = -0.011 \pm 0.014\,.
	\end{aligned}
	\label{eq:main_obs}
	\end{equation}
The largest correlation ($\approx 0.5$) involves $\Delta A_{\rm FB}$ and $\Delta \tilde F_L$.

\subsection{\boldmath $W \to \mu \nu$ and $Z \to \mu \mu$} \label{sec:Vllp}

Virtual corrections with top quarks and LQs modify couplings of gauge bosons to charged leptons, in particular to the muon~\cite{ColuccioLeskow:2016dox,Arnan:2019olv,Crivellin:2020mjs}.  Parametrizing the interactions as
\begin{align*}
&\mathcal{L}= 
\frac{g_2}{\sqrt{2}}\Lambda_{22}^{W} \left(\bar{\mu}\gamma^{\alpha}P_{L}\nu_{\mu}W_{\alpha}^{-}\right) +{\rm h.c.} \\&\;\;\;\;+ \frac{g_2}{2 c_w}\bar{\mu} \gamma^{\alpha} \left( \Lambda^V -\Lambda^A \gamma_5\right) \mu Z_{\alpha}\,,
\end{align*}
with
\begin{align*}
\Lambda^{W}_{ij}&=\delta_{ij}+\Lambda^{\rm{LQ}}_{ij} \ ,\;
\Lambda^{V,A}_{ij}=\Lambda^{V,A}_{\text{SM}} \delta_{ij} + (\Delta_{V,A}^{\rm LQ})_{ij} \ , \\ \Lambda^V_{\rm{SM}}&=-\frac{1}{2}+2s_w^2 \ , \ \Lambda^{A}_{\rm{SM}}=-\frac{1}{2} \ ,
\end{align*}
the LQ effects at $q^2=0$ (the contributions proportional to gauge-boson masses are suppressed) are given by
\begin{align}
\label{LambdaDelta}
\Lambda^{\rm LQ}_{ij}\simeq&\frac{N_c   m_t^2 }{192 \pi ^2 M^2}\Bigg[3V_{3h} \lambda _{hi}^{L*} V_{3k}^*\lambda _{kj}^{L} \left(1+2 \log \left(\frac{m_t^2}{M^2}\right)\right) \Bigg] \ ,\nonumber\\
\Delta^{\rm LQ}_{L,ij}&\simeq-V_{3l}\lambda_{li}^{L*}V_{3a}^{*}\lambda_{aj}^{L}
\frac{N_c \, m_t^2}{32 \pi^2 \, M^2} \left[ 1 + \log \left(\frac{m_t^2}{M^2} \right) \right] \ , \nonumber\\
\Delta^{\rm LQ}_{R,ij}&\simeq |\lambda^{R}_{3i}|^2 \delta_{ij} \frac{ N_c  \, {m_t^2}}{32 \pi^2 \, M^2} \left[ 1 + \log \left(\frac{m_t^2}{M^2} \right) \right] \ .
\end{align}
For illustration purposes, we only kept the dominant, $(m_t / M)^2$ terms, but we considered the full corrections (see e.g. \cite{Crivellin:2020mjs}) in the analysis.
Experimentally, the averaged modification of the $W$-$\mu \nu$ coupling extracted from $\tau \to \mu \nu \nu/\tau \to e \nu \nu$, $\pi \to \mu \nu/\pi\to e\nu$ and $K \to \mu \nu/K\to e\nu$ decays reads~\cite{Pich:2013lsa,Tanabashi:2018oca}
\begin{align}
\label{LambdaW}
{\Lambda^{W}_{22}}= 1.0018 \pm 0.0014 \ ,
\end{align}
yielding a stronger constraint than data of $W$ decays.

Concerning $Z\to\mu\mu$ the axial vector coupling is much better constrained than the vectorial one \cite{Pich:2013lsa,Tanabashi:2018oca}
\begin{align}
\label{LambdaZ}
{\Lambda^A_{22}}/{\Lambda^A_{11}} = {\Lambda^A_{22}}/{\Lambda^A_{\rm{SM}}} = 1.0002 \pm 0.0013 \ ,
\end{align}
with
$
{\Lambda^A_{22}}/{\Lambda^A_{\rm{SM}}}=1+2 \Delta_{R,22}^{\rm LQ}-2 \Delta_{L,22}^{\rm LQ}$ in our case.

\section{Phenomenology}
\label{phenomenology}

\subsection{Model-Independent Analysis}

Let us start by discussing $b\!\to\!c\ell\nu$ in the light of $\Delta A_{\mathrm{FB}}$ in a model-independent way. We consider the operators in \eq{operators}\footnote{Aspects of this model-independent analysis were already discussed in Ref.~\cite{Bobeth:2021lya}.}. Note that we did not include $O_{VR}^{\ell} = {{\bar c}}{\gamma ^\mu }{P_R}b\bar \ell {\gamma _\mu }{P_L}{\nu _\ell }$, whose Wilson coefficients, taking into account $SU(2)_L$ gauge invariance, are lepton flavour universal~\cite{Alonso:2014csa,Aebischer:2015fzz,Cata:2015lta} at the dimension-6 level and thus cannot contribute to $\Delta A_{\mathrm{FB}}$. Furthermore, we will assume that the NP effect is related to muons only, as motivated by $a_\mu$ and $b\to s\ell^+\ell^-$ data and taking into account the stringent bounds from $\mu\to e\gamma$~\cite{Crivellin:2017dsk}. The most important observables included in the fit are $R(D^{(*)})_{\mu e}$ as well as $\Delta A_{\mathrm{FB}}$, $\Delta S_3$, $\Delta F_L$ and $\Delta \tilde F_L$, summarized in eq.~(\ref{eq:main_obs}). Since a modification of the SM Wilson coefficient {$C^\ell_{VL}$} alone does not affect asymmetries, $\Delta A_{\mathrm{FB}}$ provides a unique window on scalar and/or tensor NP in $b\to c\ell\nu$ transitions. 

\begin{figure}[t]
	\centering
	\includegraphics[width=0.42\textwidth]{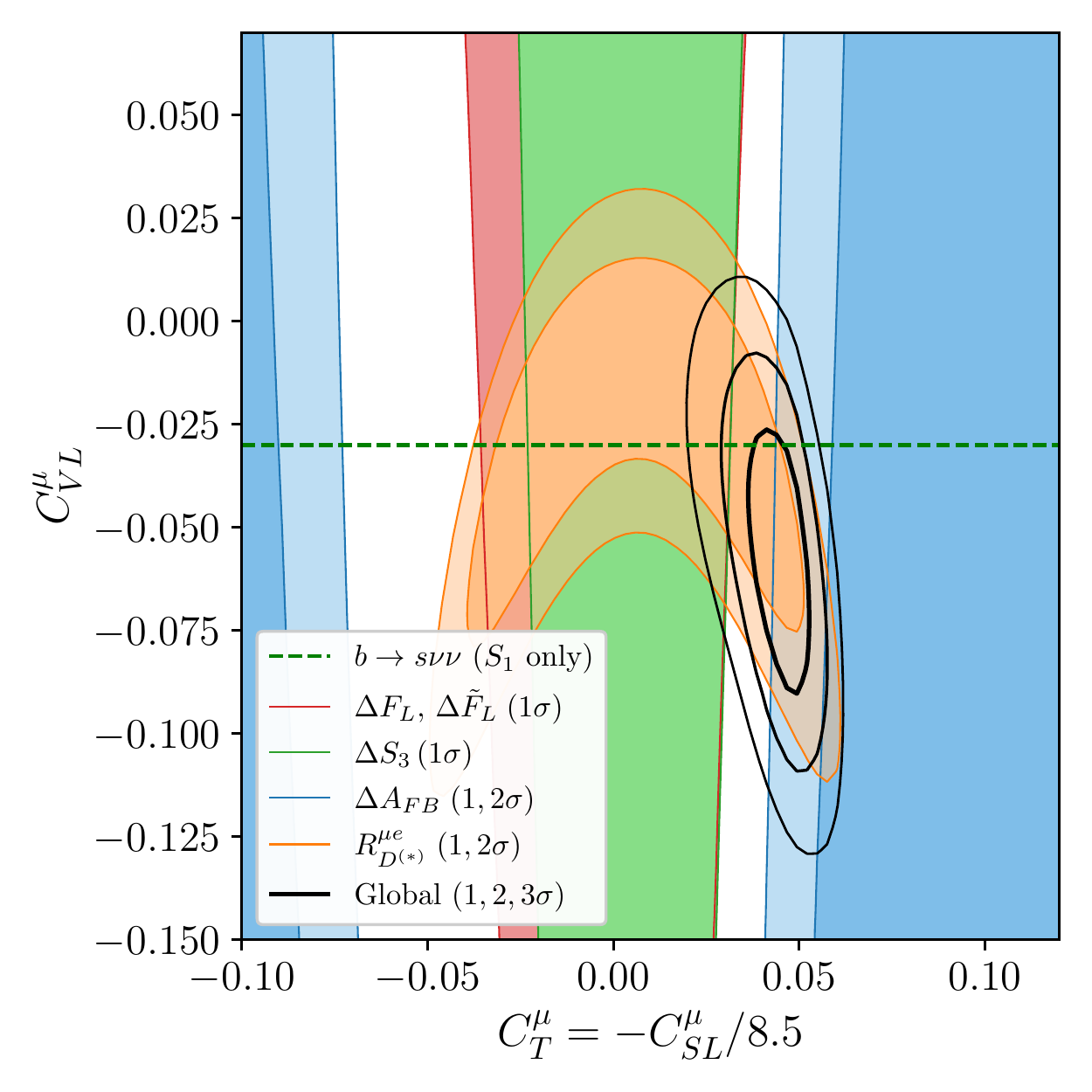}
	\caption{Preferred regions in the $C_T^\mu=-C_{SL}^\mu/8.5$ vs. $C_{VL}^\mu$ plane. One can see that a negative value of $C_{VL}^\mu$ can compensate the effect of $C_T^\mu=-C_{SL}^\mu/8.5$ in ${\rm Br}[B\to D^{(*)}\mu\nu]/{\rm Br}[B\to D^{(*)}e\nu]$ and that $\Delta A_{\mathrm{FB}}$ prefers positive values of $C_{T}^\mu$. Note that the best-fit point describes data $\approx\! 4\,\sigma$ better than the SM, even though the constraints from $\Delta S_3$, $\Delta F_L$ and $\Delta \tilde F_L$ {prefer a smaller $\Delta A_{\mathrm{FB}}$ than the current central value.}
The region below the green dashed line is excluded by $B\to K^*\nu\nu$ in case the SM is extended by the $S_1$ LQ alone.}
	\label{fig:EFTfit}
\end{figure}

Our global fit is performed using {\tt FLAVIO}~\cite{Straub:2018kue} where these observables have been pre-implemented based on the theoretical expressions of Ref.~\cite{Gratrex:2015hna} and the form factors of Refs.~\cite{Faller:2008tr, Bernlochner:2017jka, Gubernari:2018wyi,Bordone:2019vic}.
Note that for non-zero NP contributions to the Wilson coefficients the theory uncertainty increases, due to incomplete cancellations among the form-factor uncertainties. We consistently take into account this effect within {\tt FLAVIO}.
The different possible 3-dimensional fits in Table~\ref{tab:pulls} show that a non-vanishing Wilson coefficient of the tensor operator significantly improves the description of the data with respect to the SM, leading to pulls above $3\sigma$. In particular, the $C_{V_L}^{\mu}$ \& $C_{SL}^\mu = -8.5 \, C_T^\mu$ hypothesis discussed in the previous section leads to a pull of almost $4\sigma$. The fit projection to these Wilson-coefficient combinations is displayed in Fig. \ref{fig:EFTfit}. The figure shows that the best-fit region can (at 95\% CL) explain all the $\Delta$ observables while agreeing with the $b \to s \nu \nu$ and $R_{D^{(*)}}^{\mu e}$ constraints. However, note that a somewhat smaller central value of $\Delta A_{\rm FB}$ would be preferred.

\begin{table}[t]
	\renewcommand{\arraystretch}{1.4} 
	\centering
	\begin{tabular}{|l|c|c|}
		\hline
		Scenario & ~~SM pull ($\sigma$) ~~& ~~$\chi^2 /\textrm{N}_{\textrm{dof}}$~~ \\ \hline
		SM 												& -- 				& $72.9/48$ \\ 
		$C_{VL}^\mu, C_{SL}^\mu, C_{SR}^\mu, C_T^\mu  $ & $\mathbf{3.39}$ 	& $52.9/44$ \\ 
		$C_{VL}^\mu, C_{SL}^\mu, C_T^\mu  $ 			& $\mathbf{3.72}$ 	& $52.9/45$ \\ 
		$C_{VL}^\mu, C_{SL}^\mu, C_{SR}^\mu  $ 			& $2.14$ 			& $64.1/45$ \\ 
		$C_{SL}^\mu, C_{SR}^\mu, C_T^\mu  $ 			& $\mathbf{3.36}$ 	& $56.1/45$ \\ 
		$C_{VL}^\mu, C_{SL}^\mu = -8.5 \, C_T^\mu  $ 	& $\mathbf{3.94}$ 	& $54.0/46$ \\
		$C_{VL}^\mu, C_{SL}^\mu = 8.5 \, C_T^\mu $ 		& $2.42$ 			& $64.5/46$ \\ \hline
	\end{tabular}
\caption{$\chi^2/\textrm{N}_{\textrm{dof}}$ and pulls (w.r.t.~the SM) of the global fit for different scenarios. Fits with a non-zero Wilson coefficient for the tensor lead to pulls above $3\sigma$. 
}
\label{tab:pulls}
\end{table}

Therefore, restricting ourselves to tree-level renormalizable SM extensions, only two scalar LQs may potentially explain the $\Delta A_{\mathrm{FB}}$ without violating $R(D^{(*)})_{\mu e}$. In fact, among the two LQs that generate Wilson coefficients of the tensor operator, both in combination with a $C_{SL(R)}^\ell$ contribution, the $SU(2)_L$ singlet (doublet) $S_1$ ($S_2$) results in $C_{SL}^\ell=-(+)4C_T^\ell$ at the matching scale and only $S_1$ generates $C_{VL}^\ell$. Interestingly, the $C_{VL}^\mu$, $C_{SL}^\mu=-4C_T^\mu$ scenario (corresponding to $C_{SL}^\mu\approx-8.5\, C_T^\mu$ at the $B$ scale) is also the best-performing hypothesis in Table~\ref{tab:pulls}, with a pull of $3.9\sigma$ with respect to the SM hypothesis. The best-fit point for $C_T$ is 0.044 and the $1\sigma$ interval is $[0.036, 0.062]$. Hence this model-independent analysis gives strong support in favour of the $S_1$ LQ.

\subsection{Leptoquark Analysis}

Let us now reconsider the fit of Fig.~\ref{fig:EFTfit} in the context of the singlet LQ $S_1$. Here, the relevant difference compared to the model-independent setup is that $C_{VL}^\mu$ is constrained by $b\to s\nu\nu$. In particular, if $\lambda^L_{22}$ is zero, such that $C^\mu_{VL}$ would only be induced via CKM rotations from $\lambda^L_{32}$, $C_{VL}^\mu$ would be positive. In order to obtain a negative value for $C_{VL}^\mu$, $\lambda^L_{22}$ is required. However, together with $\lambda^L_{32}$ this induces an effect in $B\to K^*\nu\nu$ which bounds $C_{VL}^\mu \gtrsim -0.03$. Note that the bound from $B\to K^*\nu\nu$ could be avoided by adding the triplet scalar LQ $S_3$~\cite{Crivellin:2017zlb}.

\begin{figure}[t]
	\centering
	\includegraphics[width=0.48\textwidth]{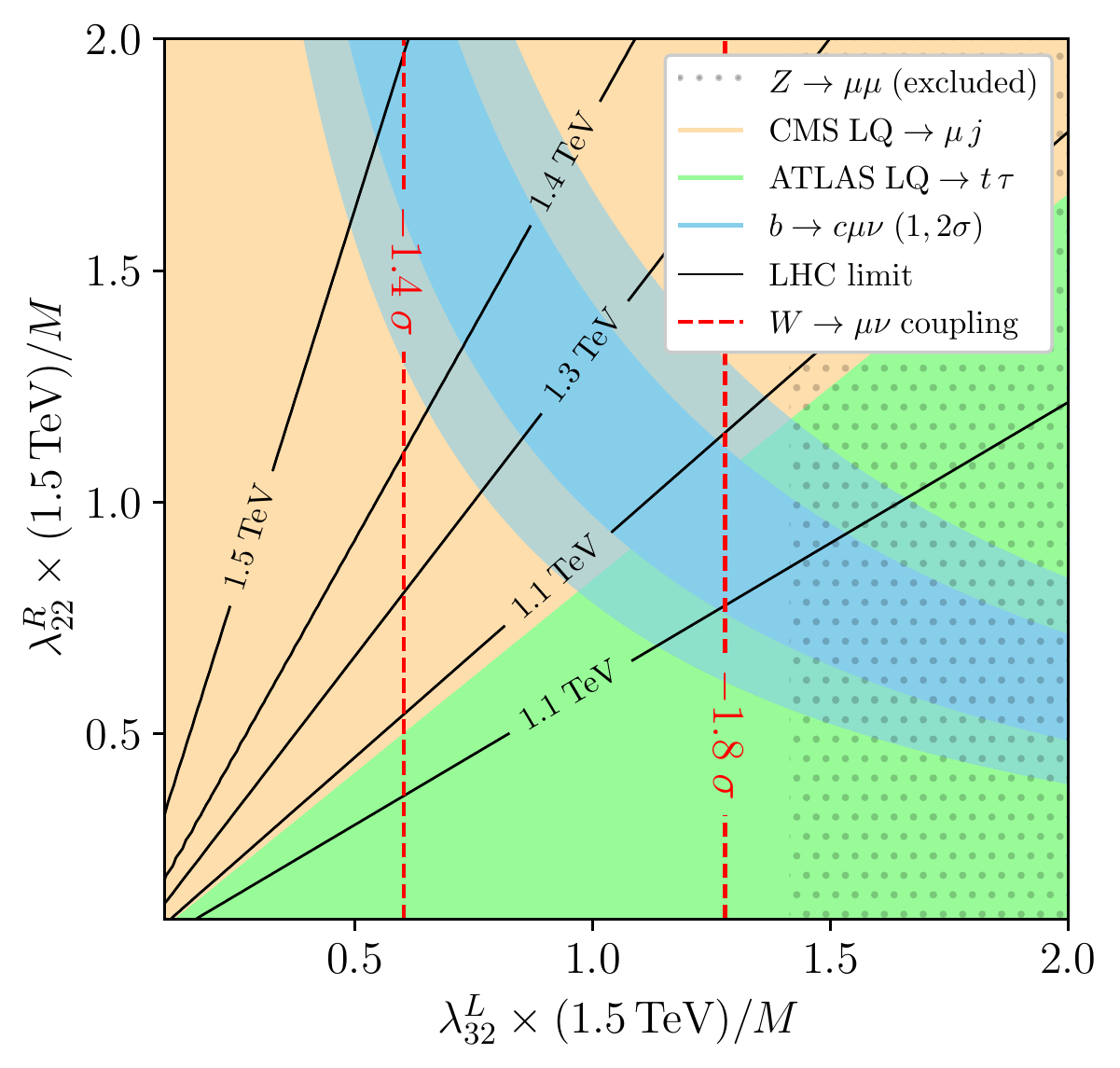}
	\caption{Preferred and excluded regions in the $\lambda^L_{32}$--$\lambda^R_{22}$ plane for $M=1.5\,$TeV for the $S_1$ LQ profiling over $C_{VL}^\mu$. The contour lines show the lower limits on the mass $M$ from ATLAS and CMS searches for third and second generation LQs via the final states $t\tau$ and $j\mu$, respectively. The dotted region is excluded by LEP measurements of $Z\to\mu\mu$ and vertical dashed contour lines show the tension in the $W$-$\mu \nu$ coupling.}
	\label{fig:LQfit}
\end{figure}

Profiling over $C_{VL}^\mu$ we show the preferred regions in the $\lambda^L_{32}$--$\lambda^R_{22}$ plane for $M=1.5\,$TeV in Fig.~\ref{fig:LQfit}, where we assume real couplings. Here the contour lines show the current lower limit on the mass $M$ of $S_1$ from LHC searches.
In the green region the bounds from searches for pair production of LQs with top-tau ($t\tau$) final states of ATLAS~\cite{Aaboud:2019bye} dominate, while in the orange region the CMS bounds~\cite{Sirunyan:2018ryt} on LQ decaying to jet and muons ($j\mu$) are more stringent (see also \cite{Angelescu:2018tyl,Angelescu:2021lln}). The grey hatched region is excluded by LEP measurements of $Z\to\mu\mu$. Note that this bound is also sensitive to a possible extension of the $S_1$ model and should therefore only be considered as an estimate. Finally, Fig.~\ref{fig:LQfit} indicates that the preferred region shows some tension with the bounds on $W$-$\mu\nu$ coupling \footnote{In fact, the effect would increase the tension in the Cabibbo angle anomaly~\cite{Crivellin:2020lzu}. However, here LQs could also contribute at tree level alleviating this tension~\cite{Crivellin:2021egp}.}. However, also this bound is sensitive to additions of new particles, e.g.~$S_3$ of the singlet-triplet model~\cite{Crivellin:2017zlb}, where a constructive effect in $W\to\mu\nu$ is generated~\cite{Crivellin:2020mjs}.

\section{Conclusions and Outlook}
\label{conclusions}

In addition to the existing intriguing hints for the violation of lepton flavour universality, Ref.~\cite{Bobeth:2021lya} recently pointed out that, in the light of the data in Ref.~\cite{Waheed:2018djm} {and pending certain reservations}, the forward-backward asymmetry of $ B\to D^{*}\mu\bar \nu$ vs $ B\to D^{*}e\bar \nu$ ($\Delta A_{\rm FB}$) {shows a tension} with the SM prediction by $\approx\!4\,\sigma$. In this article we found that, in order to significantly improve the global fit w.r.t. the SM, one needs NP in the tensor operator. As, within renomalizable models at tree level, non-vanishing Wilson coefficients of this operator can only be induced by scalar LQs \cite{Crivellin:2017dsk,deBlas:2017xtg}, we investigate the two model-independent scenarios motivated by them, finding that only the scenario $C_{VL}, C_{SL}=-4C_T$, related to the $SU(2)_L$-singlet scalar LQ $S_1$, improves significantly ($\approx 4\,\sigma$) the description of data.

\begin{figure}[t]
	\centering
	\includegraphics[width=0.48\textwidth]{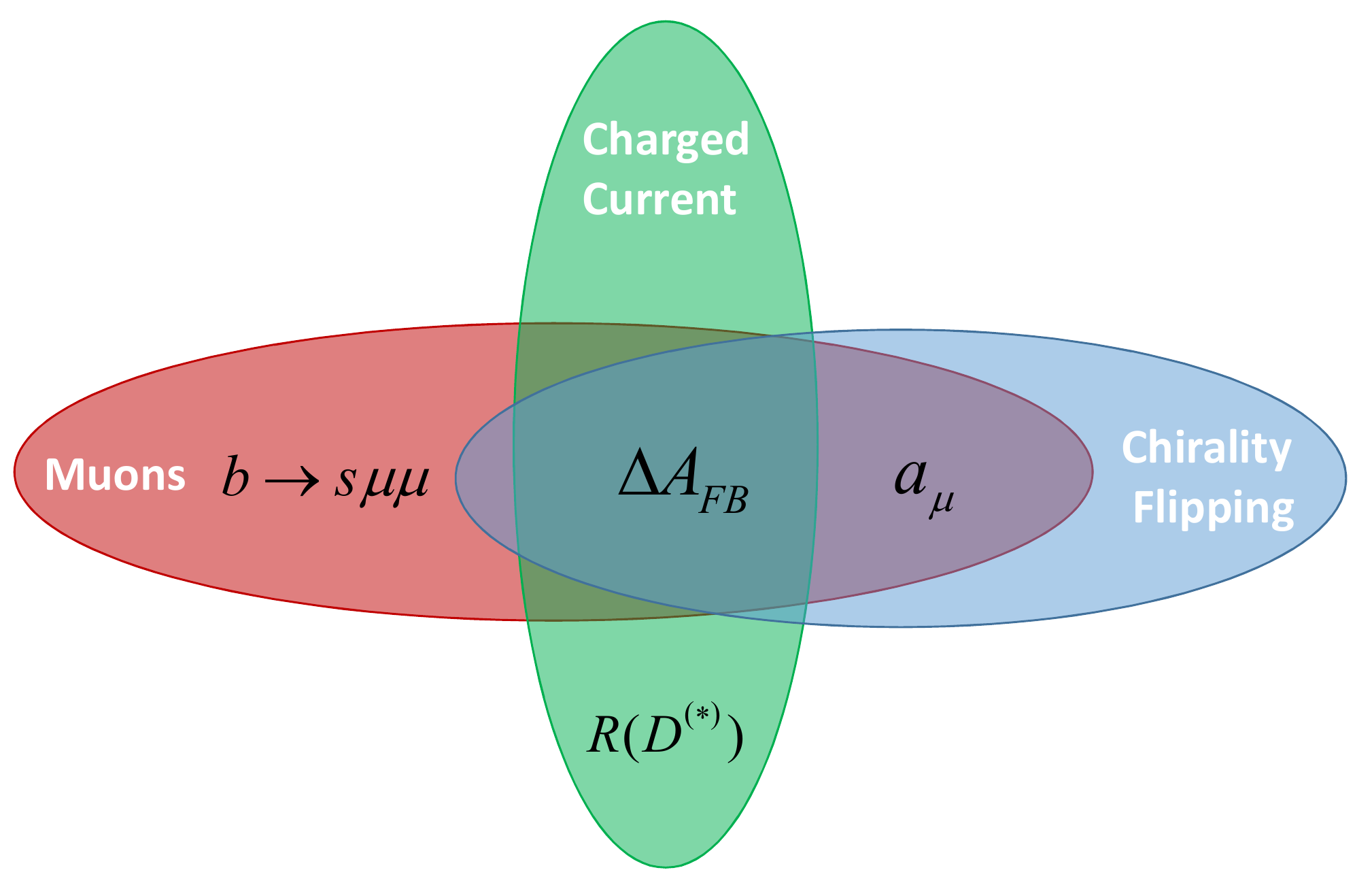}
	\caption{Relations among the different hints for LFUV. As one can see, $\Delta A_{\rm FB}$ lies in the center of the graph, as it is a charged-current process, involves muons and its explanation requires chirality flipping operators. This justifies the special role of this observable.}
	\label{fig:relations}
\end{figure}

Our findings call for a reanalysis of the Belle data \cite{Waheed:2018djm}, focusing on the theoretically cleaner $\Delta$ observables discussed in \cite{Bobeth:2021lya}. In the light of our results---the circumstantial evidence for scalar LQs, in particular $S_1$---a confirmation of the $\Delta A_{\rm FB}$ discrepancy would not only represent the first evidence of muon-related NP in $b \to c$ transitions, but would also allow for a natural connection with the all the other muonic anomalies, including $b \to s \mu\mu$ data and $(g - 2)_\mu$. In fact, as can be seen in Fig.~\ref{fig:relations}, $\Delta A_{\rm FB}$ combines several key properties of the other anomalies.
\medskip

\begin{acknowledgments}
DG warmly acknowledges the authors of Ref.~\cite{Bobeth:2021lya}, in particular Christoph Bobeth and Danny van Dyk, for insights, plots, and useful remarks on the draft. We also thank Peter Stangl for comments as well as for constant feedback about {\tt FLAVIO}. The work of AnC is supported by a Professorship Grant (PP00P2\_176884) of the Swiss National Science Foundation. The work of AlC, SG and DG is supported by the ANR under contract n. 202650 (PRC `GammaRare').	
\end{acknowledgments}

\bibliography{bibliography}

\end{document}